\documentstyle[12pt]{paper}
\author{S.\ M.\ Troshin,
 N.\ E.\ Tyurin\\
Institute for High Energy Physics\\
142284 Protvino, Moscow Region, Russia}
\title{ Beyond the black disk limit \rm}
\date{}
\begin{document}
\maketitle
\begin{abstract}
We consider consequences of violation of the black disk
limit possibly revealed by the new CDF measurements of
 the total, elastic and diffractive cross--sections.
\\[1ex]

\end{abstract}

The new CDF data \cite{cdf} on the measurements of the total, elastic and
diffractive $p\bar{p}$--scattering cross--sections at Tevatron--Collider
($\sqrt{s}=1.8$ TeV) reveal interesting and unexpected results:
\begin{itemize}
\item
large total cross--section $\sigma _{tot}=80.6\pm 2.3$ mb which is
consistent with $\ln^2s$--rise of $\sigma _{tot}$;
\item
large elastic cross--section $\sigma _{el}=20.0\pm 0.9$ mb and large
 ratio of elastic to total cross--section
$\sigma _{el}/\sigma _{tot}=0.248\pm0.005$
which show the greater the energy the larger both absolute and relative
probabilities of elastic collisions;
\item
scattering amplitude in the impact parameter representation has
the value Im$f(s,b=0)=0.50\pm0.01$.
\end{itemize}
With the data on the slope parameter of diffraction cone and
single diffractive cross--section also measured by CDF, these
results point out: the black disk limit is almost reached
at small impact parameters, the scattering amplitude at $b=0$ is
probably beyond this limit, the Pumplin bound for
diffractive cross--section \cite{pum} is violated at
$b=0$ \cite{ttpl}, \cite{fra}.
Of course, these conclusions should be taken with certain
precautions since the E710 gives different figures for
the cross--sections \cite{roy}.

However, the possibility that the scattering amplitude exceed the black
disk limit at $b=0$ seems worth to be discussed.
First of all, let us remind that the unitary equation for the
scattering amplitude
\begin{equation}
\mbox{Im} f(s,b)=|f(s,b)|^2+\eta (s,b),\quad \eta (s,b)=\sum_n \sigma _n(s,b),
\label{u}
\end{equation}
imply the constraint $|f(s,b)|\leq 1$ while the black disk limit
assumes that $|f(s,b)|\leq 1/2$. The equality $|f(s,b)|=1/2$
corresponds to the maximal absorption at given values of $s$ and
$b$.
The Pumplin bound for the diffractive cross--section
\begin{equation}
\sigma _{diff}(s,b)\leq \frac{1}{2}\sigma _{tot}(s,b)-\sigma _{el}(s,b)
\label{p}
\end{equation}
is also based on the assumption that the diffractive
eigenamplitudes in the Good--Walker picture \cite{gw} do not exceed
the black disk limit.
Indeed this limit is a priori chosen by the use of the
imaginary eikonal $\Omega =i\chi $ when the amplitude  is
written in the following form to ensure unitarity:
\begin{equation}
f(s,b)=\frac{i}{2}(1-\exp[i\Omega (s,b)]). \label{e}
\end{equation}

There is another possibility to provide the direct channel
unitarity. It is based on the following representation for
an amplitude \cite{log}:
\begin{equation}
f(s,b)=U(s,b)[1-iU(s,b)]^{-1}. \label{um}
\end{equation}
In the latter case the inelastic channel contribution takes the
following form
\begin{equation}
\eta (s,b)=\mbox{Im} U(s,b)|1-iU(s,b)|^{-2}. \label{in}
\end{equation}
Eq. \ref{in} ensures $s$--channel unitarity provided that
$\mbox{Im}U(s,b)\geq 0$.

Usually eikonal or $U$--matrix are considered to be input
dynamical quantities similar to the Born term (however, the
expansion in these terms is not always possible) and there are
number of dynamical models used for the construction of explicit
form of eikonal or $U$--matrix. The most of QCD--inspired,
Regge--type and geometrical models lead to the parameterization of
the above quantities in the form
\[
\Omega ,U(s,b)\propto is^\lambda e^{-b^2/a(s)},\quad a(s)\sim \ln s
\]
or
\[
\Omega ,U(s,b)\propto is^\lambda e^{-\mu b},
\] where $\mu $ is a constant. The latter form respects
analytical properties of the scattering amplitude in the complex
$\cos \theta $--plane.

These parameterizations of $\Omega $ and $U$ and unitarization
methods lead to growth of the total and elastic cross--sections
\[
\sigma _{tot}(s)\sim \sigma _{el}(s)\sim \ln^2 s.
\]

However, the above two unitarization
methods are different, e.g. they give different behavior of the ratio
of elastic to total cross--sections:
\[
\sigma _{el}(s)/\sigma _{tot}(s)\rightarrow
1/2
\]
 for eikonal unitarization and
\[
\sigma _{el}(s)/\sigma _{tot}(s)\rightarrow 1
\]
 for the $U$--matrix unitarization. These asymptotical regimes reflect
the limitations for the scattering amplitude $|f(s,b)|\leq 1/2$
and $|f(s,b)|\leq 1$ which are imposed by the imaginary eikonal
and $U$--matrix respectively.

Therefore the $U$--matrix unitarization implies the two modes in hadron
scattering: shadow and antishadow, while imaginary eikonal
corresponds to the shadow scattering only.

 To clarify the difference
between the two scattering modes we
consider pure imaginary amplitude. The unitary equation has two
solutions
\begin{equation}
f(s,b)=\frac{i}{2}[1\pm \sqrt{1-4\eta (s,b)}].\label{*}
\end{equation}
In the case of shadow scattering an elastic
amplitude increases with increase of the contribution of inelastic
channels. This mode corresponds to the choice of minus
 sign.
In the antishadow mode (sign $+$) the
amplitude increases with decrease of the inelastic channel
contribution. The shadow scattering mode is often considered as  only
possible one, however, it should be noted that these two solutions
have equal meaning and at high energies the other mode could be
 realized.

Let us consider such a transition from the shadow to antishadow
scattering mode.

In the framework of
the $U$--matrix unitarization scheme the inelastic overlap function
$\eta (s,b=0)$ increases with energy. It gets maximum
$\eta (s,b=0)=1/4$ at some energy $s=s_0$ and beyond that point the
transition to the antishadow mode occurs when Im$f(s,b=0)>1/2$
and  $\eta (s,b=0)<1/4$.
Note that value of $\sqrt{s_0}=2$ TeV was predicted in Refs. \cite{tt}.

The function $\eta (s,b)$ becomes peripheral when
energy grows. At $s>s_0$ the maximum of $\eta (s,b)$ is
reached at $b=R(s)$ where $R(s)$ is the interaction radius
\cite{tu}. This
picture corresponds to the antishadow scattering at $b<R(s)$ and to
the shadow scattering at $b>R(s)$. At $b=R(s)$ the maximal absorption
takes place.

The transition to the antishadow mode in the $U$--matrix
method naturally occurs when $U(s,b=0)$ becomes greater than unity and in
this case one cannot expand the amplitude into the series over
 $U$. Of course, such a transition to the antishadow mode, in
principle, could be realized in the eikonal approach also. However, this
transition in particular imply that real part of eikonal $\Omega $ gain abrupt
increase  equal to $\pi $ at $s=s_0$. The commonly accepted models
for the eikonals do not foresee such a behavior.

It is to be  emphasized  that the CDF data
indicate that $\sqrt{s}=1.8$ TeV is in some sense threshold
energy, namely, the amplitude pergaps goes beyond the black disk
limit at zero impact parameter  and the antishadow scattering mode
starts to develop in the central hadron collisions.

The  hadron scattering
picture may be described  as transition from grey to black
disk and then to black ring with energy growth.
The black ring picture is in complete agreement
with the unitarity.

We would like to thank M. M. Islam, A. D. Krisch and V. A. Petrov
for useful comments.


\small
\begin{thebibliography}{99}
\bibitem{cdf}
CDF Collaboration, talk presented at the Vth Blois Conference on
Elastic and Diffractive Scattering, Providence, 8--12 June, 1993.
\bibitem{pum}
J. Pumplin, \it Phys. Rev., \rm \bf D8\rm (1973) 2899.
\bibitem{ttpl}
S. M. Troshin and N. E. Tyurin, \it Phys. Lett. \rm \bf 208 B \rm
(1988) 517;
\bibitem{fra}
L. L. Frankfurt, talk presented at the Vth Blois Conference on
Elastic and Diffractive Scattering, Providence, 8--12 June, 1993.
\bibitem{roy}
R. Rubinshtein, talk presented at the Vth Blois Conference on
Elastic and Diffractive Scattering, Providence, 8--12 June, 1993.
\bibitem{gw}
M. M. Good and W. D. Walker, \it Phys. Rev. \rm \bf 120 \rm (1960)
1857.
\bibitem{log}
A. A. Logunov, V. I. Savrin, N. E. Tyurin and O. A. Khrustalev, \it  Teor.
Mat. Fiz. \rm \bf  6 \rm(1971) 157.
\bibitem{tt}
S. M. Troshin, N. E. Tyurin and  O.  P.  Yuschenko, \it  Nuovo  Cim.
\rm \bf 91A\rm (1986) 23; S. M. Troshin and N. E. Tyurin, paper
presented at XXIII International Conference on High Energy
Physics, Berkeley, 1986; Preprint IHEP 86--232, Serpukhov, 1986.
\bibitem{tu}
N. E. Tyurin, \it Nucl. Phys. B (Proc. Suppl.) \rm \bf 25B\rm (1992)
91.
\end{thebibliography}
\end{document}